\pdfoutput=1
\documentclass[usenatbib]{mn2e}
\usepackage{siunitx}
\usepackage{graphicx}
\usepackage{latexsym}
\usepackage{aas_macros}
\usepackage{amsfonts,amsmath,amssymb}
\usepackage{url}
\usepackage[utf8]{inputenc}
\usepackage{fancyref}
\usepackage{subfloat}
\usepackage{verbatim}

\title[SDSS J1057+2759: a period-bounce CV]{SDSS J105754.25+275947.5: a period-bounce eclipsing cataclysmic variable with the lowest-mass donor yet measured}

\author[M.\,J.\ McAllister et al.]{M.\,J.\ McAllister$^{1}$, S.\,P.\ Littlefair$^{1}$, V.\,S.\ Dhillon$^{1,2}$, T.\,R.\ Marsh$^{3}$, B.\,T.\ G\"ansicke$^{3}$,
\newauthor J.\ Bochinski$^{4}$, M.\,C.\,P.\ Bours$^{3,5}$, E.\ Breedt$^{3}$, L.\,K.\ Hardy$^{1}$, J.\,J.\ Hermes$^{6}$, 
\newauthor S.\ Kengkriangkrai$^{7}$, P.\ Kerry$^{1}$, S.\,G.\ Parsons$^{1}$, S.\ Rattanasoon$^{1,7}$\\
$^{1}$Department of Physics and Astronomy, University of Sheffield, Sheffield, S3 7RH, UK\\
$^{2}$Instituto de Astrofisica de Canarias, E-38205 La Laguna, Tenerife, Spain\\
$^{3}$Department of Physics, University of Warwick, Coventry, CV4 7AL, UK\\
$^{4}$Department of Physical Sciences, The Open University, Milton Keynes, MK7 6AA, UK\\
$^{5}$Departmento de F\'isica y Astronom\'ia, Universidad de Valpara\'iso, Avenida Gran Bretana 1111, Valpara\'iso, 2360102, Chile\\
$^{6}$Hubble Fellow - Department of Physics, University of North Carolina, Chapel Hill, NC 27599, USA\\
$^{7}$National Astronomical Research Institute of Thailand, 191 Siriphanich Building, Huay Kaew Road, Chiang Mai 50200, Thailand}

\begin{document}



\maketitle

\label{firstpage}

\begin{abstract}

\noindent We present high-speed, multicolour photometry of the faint, eclipsing cataclysmic variable (CV) SDSS J105754.25+275947.5. The light from this system is dominated by the white dwarf. Nonetheless, averaging many eclipses reveals additional features from the eclipse of the bright spot. This enables the fitting of a parameterised eclipse model to these average light curves, allowing the precise measurement of system parameters. We find a mass ratio of $q\,=\,0.0546\,\pm\,0.0020$ and inclination $i\,=\,85.74\,\pm\,0.21\,^{\circ}$. The white dwarf and donor masses were found to be $M_{\mathrm{w}}\,=\,0.800\,\pm\,0.015\,$M$_{\odot}$ and $M_{\mathrm{d}}\,=\,0.0436\,\pm\,0.0020\,$M$_{\odot}$, respectively. A temperature $T_{\mathrm{w}}$\,=\,13300\,$\pm$\,1100\,K and distance $d\,=\,367\,\pm\,26$\,pc of the white dwarf were estimated through fitting model atmosphere predictions to multicolour fluxes. The mass of the white dwarf in SDSS 105754.25+275947.5 is close to the average for CV white dwarfs, while the donor has the lowest mass yet measured in an eclipsing CV. A low-mass donor and an orbital period (90.44\,min) significantly longer than the period minimum strongly suggest that this is a bona fide period-bounce system, although formation from a white dwarf/brown dwarf binary cannot be ruled out. Very few period-minimum/period-bounce systems with precise system parameters are currently known, and as a consequence the evolution of CVs in this regime is not yet fully understood.


\end{abstract}


\bibliographystyle{mn2e_fixed2}

\section{Introduction} 
\label{sec:introduction}

Cataclysmic variable stars (CVs) are close, interacting binary systems containing a white dwarf primary star and a low-mass, Roche-lobe filling secondary star. Material from the secondary (donor) star is transferred to the white dwarf, but is not immediately accreted in those systems with a low magnetic field white dwarf. Instead, an accretion disc forms in order for angular momentum to be conserved. An area of increased luminosity is present at the point where the stream of transferred material makes contact with the disc, and is termed the bright spot. For a general review of CVs, see \cite{warner95} and \cite{hellier01}.

For systems with inclinations greater than approximately 80$^{\circ}$ to our line of sight, the donor star can eclipse all other system components. Eclipses of the individual components -- white dwarf, bright spot and accretion disc -- create a complex eclipse shape. These individual eclipses occur in quick succession, and therefore high-time resolution observations are required in order to separate them from each other. High-time resolution also allows the timings of white dwarf and bright spot eclipses to be precisely measured, which can be used to derive accurate system parameters \citep{wood86}.

Steady mass transfer from donor to white dwarf is possible in CVs due to sustained angular momentum loss from the system. The gradual loss of angular momentum causes the donor star's radius -- and therefore the separation and orbital period of the system --  to decrease over time. During this process the donor star's thermal time-scale increases at a faster rate than its mass-loss time-scale, which has the effect of driving it further away from thermal equilibrium. Around the point where the donor star becomes substellar, it is sufficiently far from thermal equilibrium for it to no longer shrink in response to angular momentum loss. In fact, the (now degenerate) donor star's radius actually increases with further losses, resulting in the system separation and orbital period also increasing (e.g. \citealt{rappaport82}; \citealt{knigge11}).

A consequence of CV evolution is therefore a minimum orbital period that systems reach before heading back towards longer periods. The orbital period minimum is observed to occur at an orbital period of 81.8\,$\pm$\,0.9\,min \citep{knigge11}, consistent with an accumulation of systems found at 82.4\,$\pm$\,0.7\,min \citep{gaensicke09} known as the `period spike' and expected to coincide with the period minimum. CVs evolving back towards longer periods are referred to as period-bounce systems or `period bouncers'.

When considering period bouncers as a fraction of the total CV population, there is a serious discrepancy between prediction and observation. Evolutionary models predict $\sim$\,40-70\% of the total CV population to be period bouncers \citep{kolb93,goliasch15}. In contrast, for many years there was a distinct lack of direct evidence for any substellar donors within CVs \citep{littlefair03}, and it wasn't until a decade ago that the first direct detection was claimed by \cite{littlefair06}. A rough estimate of $\sim$\,15\% for the fraction of period bouncers was made from a small sample of eclipsing CVs by \cite{savoury11}. Two characteristics of period-bounce CVs are a faint quiescent magnitude and a long outburst recurrence time \citep{patterson11}, which may have resulted in an under-sampling of the population. However, the identification of CVs from the Sloan Digital Sky Survey (SDSS; \citealt{york00}) -- which make up the majority of Savoury et al.'s sample -- should not be significantly affected by either, due to being reasonably complete down to $g\sim19$ mag and selection from spectral analysis \citep{gaensicke09}. We therefore expect to find a substantial population of period-bounce systems in the SDSS sample.

One such object discovered by the SDSS is SDSS J105754.25+275947.5 (hereafter SDSS 1057). A faint system at $g'\simeq19.5$, it was identified as a CV by \cite{szkody09}. The SDSS spectrum for this system is dominated by the white dwarf, and also shows double-peaked Balmer emission lines -- characteristic of a high-inclination binary. \cite{southworth15} confirmed SDSS 1057 to be an eclipsing CV after finding short and deep eclipses with low-time-resolution photometry. These light curves also appear flat outside of eclipse with no obvious orbital hump before eclipses, hinting at  a faint bright spot feature and therefore low accretion rate. From their photometry, \cite{southworth15} measure SDSS 1057's orbital period to be 90.44\,min. Due to a low accretion rate and no sign of a secondary star in its spectrum, \cite{southworth15} highlight SDSS 1057 as a good candidate for a period-bounce system.

In this paper, we present high-time resolution ULTRACAM and ULTRASPEC eclipse light curves of SDSS 1057, which we average and model in order to obtain precise system parameters. The observations are described in Section~\ref{sec:observations}, the results displayed in Section~\ref{sec:results}, and an analysis of these results is presented in Section~\ref{sec:discussion}.

\section{Observations}
\label{sec:observations}

SDSS 1057 was observed a total of 12 times from Apr 2012 - Jun 2015 with the high-speed cameras ULTRACAM \citep{dhillon07} and ULTRASPEC \citep{dhillon14}. Half of these observations are from ULTRACAM on the 4.2\,m William Herschel Telescope (WHT), La Palma, with the other half from ULTRASPEC on the 2.4\,m Thai National Telescope (TNT), Thailand. Eclipses were observed simultaneously in the SDSS $u' g' r'$ filters with ULTRACAM and in a Schott KG5 filter with ULTRASPEC. The Schott KG5 filter is a broad filter, covering approximately $u' + g' + r'$. A complete journal of observations is shown in Table~\ref{table:obs}.

Data reduction was carried out using the ULTRACAM pipeline reduction software (see \citealt{dhillon07}). A nearby, photometrically stable comparison star was used to correct for any transparency variations during observations.

The standard stars Feige 34 (observed on 29 Apr 2012), G162-66 (25 Apr 2012) and HD 121968 (21 and 23 Jun 2015) were used to transform the photometry into the $u' g' r' i' z'$ standard system \citep{smith02}. The KG5 filter was calibrated using a similar method to \cite{bell12}; see \citeauthor{hardy16} (2016, in press) for a full description of the calibration process. A KG5 magnitude was calculated for the SDSS standard star GJ 745A (01 Mar 2015), and used to find a target flux in the KG5 band. 

For observations at the WHT, photometry was corrected for extinction using the typical $r'$-band extinction for good quality, dust free nights from the Carlsberg Meridian Telescope\footnote{\url{http://www.ast.cam.ac.uk/ioa/research/cmt/camc_extinction.html}}, and subsequently converted into $u'$ and $g'$ bands using the information provided in La Palma Technical Note 31\footnote{\url{http://www.ing.iac.es/Astronomy/observing/manuals/ps/tech_notes/tn031.pdf}}. At the TNT, photometry was corrected using extinction measurements obtained during the commissioning phase (Nov 2013) of ULTRASPEC  \citep{dhillon14}.

\begin{table*}
\begin{center}
\begin{tabular}{lcccclccccc}
\hline
Date & Start & End & Instrument & Filter(s) & $T_{\mathrm{mid}}$ & $T_{\mathrm{exp}}$ & $N_{u'}$ & $N_{\mathrm{exp}}$ & Seeing & Airmass \\
& Phase & Phase & Setup && (HMJD) & (secs) &&& (arcsecs) & \\ \hline
2012 Apr 28 & -0.581 & 0.149 & WHT+UCAM & $u'g'r'$ & 56046.002399(12) & 4.021 & 3 & 981 & 1.1-3.0 & 1.06-1.20 \\
2012 Apr 29 & 14.761 & 15.228 & WHT+UCAM & $u'g'r'$ & 56046.944270(12) & 4.021 & 3 & 628 & 1.1-2.0 & 1.01-1.05 \\
2013 Dec 30 & 9734.442 & 9735.320 & WHT+UCAM & $u'g'r'$ & 56657.28205(3) & 4.021 & 3 & 1178 & 1.0-1.6 & 1.00-1.08 \\
2014 Jan 25 & 10140.684 & 10141.064 & TNT+USPEC & KG5 & 56682.775595(12) & 4.877 & -- & 422 & 1.4-2.7 & 1.08-1.15 \\
2014 Nov 28 & 15030.775 & 15031.138 & TNT+USPEC & KG5 & 56989.82829(3) & 3.945 & -- & 498 & 1.3-2.5 & 1.66-2.01 \\
2014 Nov 29 & 15047.842 & 15048.143 & TNT+USPEC & KG5 & 56990.89570(3) & 4.945 & -- & 331 & 0.9-1.4 & 1.16-1.25 \\
2015 Feb 24 & 16432.781 & 16433.281 & TNT+USPEC & KG5 & 57077.862577(12) & 11.852 & -- & 230 & 1.4-2.1 & 1.16-1.32 \\
2015 Feb 25 & 16448.793 & 16449.169 & TNT+USPEC & KG5 & 57078.867265(12) & 11.946 & -- & 172 & 1.9-2.8 & 1.19-1.32 \\
2015 Mar 01 & 16512.899 & 16513.138 & TNT+USPEC & KG5 & 57082.885950(12) & 11.852 & -- & 111 & 1.4-1.8 & 1.41-1.54 \\
2015 Jun 21 & 18296.610 & 18297.182 & WHT+UCAM & $u'g'r'$ & 57194.906824(12) & 4.021 & 3 & 769 & 1.2-2.1 & 1.30-1.59 \\
2015 Jun 22 & 18312.830 & 18313.171 & WHT+UCAM & $u'g'r'$ & 57195.911476(12) & 4.021 & 3 & 460 & 1.2-2.3 & 1.45-1.66 \\
2015 Jun 23 & 18328.821 & 18329.130 & WHT+UCAM & $u'g'r'$ & 57196.916157(12) & 4.021 & 3 & 416 & 1.1-2.0 & 1.51-1.72 \\
\hline
\end{tabular}
\caption{\label{table:obs}Journal of observations. The dead-time between exposures was 0.025\,s and 0.015\,s for ULTRACAM (UCAM) and ULTRASPEC (USPEC) observations, respectively. The relative timestamping accuracy is of order 10\,$\mu$s, while the absolute GPS timestamp on each data point is accurate to $<$\,1\,ms. $T_{\mathrm{mid}}$ represents the mid-eclipse time, $T_{\mathrm{exp}}$ the exposure time and $N_{\mathrm{exp}}$ the number of exposures. $N_{u'}$ indicates the number of $u'$ band frames which were co-added on-chip to reduce the impact of readout noise.}
\end{center}
\end{table*}

\section{Results}
\label{sec:results}

\subsection{Orbital ephemeris}
\label{subsec:orbeph}

Mid-eclipse times ($T_{\mathrm{mid}}$) were determined assuming that the white dwarf eclipse is symmetric around phase zero: $T_{\mathrm{mid}} = (T_{\mathrm{wi}} + T_{\mathrm{we}})/2$, where $T_{\mathrm{wi}}$ and $T_{\mathrm{we}}$ are the times of white dwarf mid-ingress and mid-egress, respectively. $T_{\mathrm{wi}}$ and $T_{\mathrm{we}}$ were determined by locating the times of minimum and maximum in the smoothed light curve derivative. There were no significant deviations from linearity in the $T_{\mathrm{mid}}$ values and the $T_{\mathrm{mid}}$ errors (see Table~\ref{table:obs}) were adjusted to give $\chi^{2}$ = 1 with respect to a linear fit.

All eclipses were used to determine the following ephemeris:
\\
\\
\begin{equation}
HMJD = 56046.002389(8) + 0.0627919557(6) E.
\label{eq:ephem}
\end{equation}
\\
\\
This ephemeris was used to phase-fold the data for the analysis that follows.

\begin{figure}
\begin{center}
\includegraphics[width=1.0\columnwidth,trim=30 20 20 20]{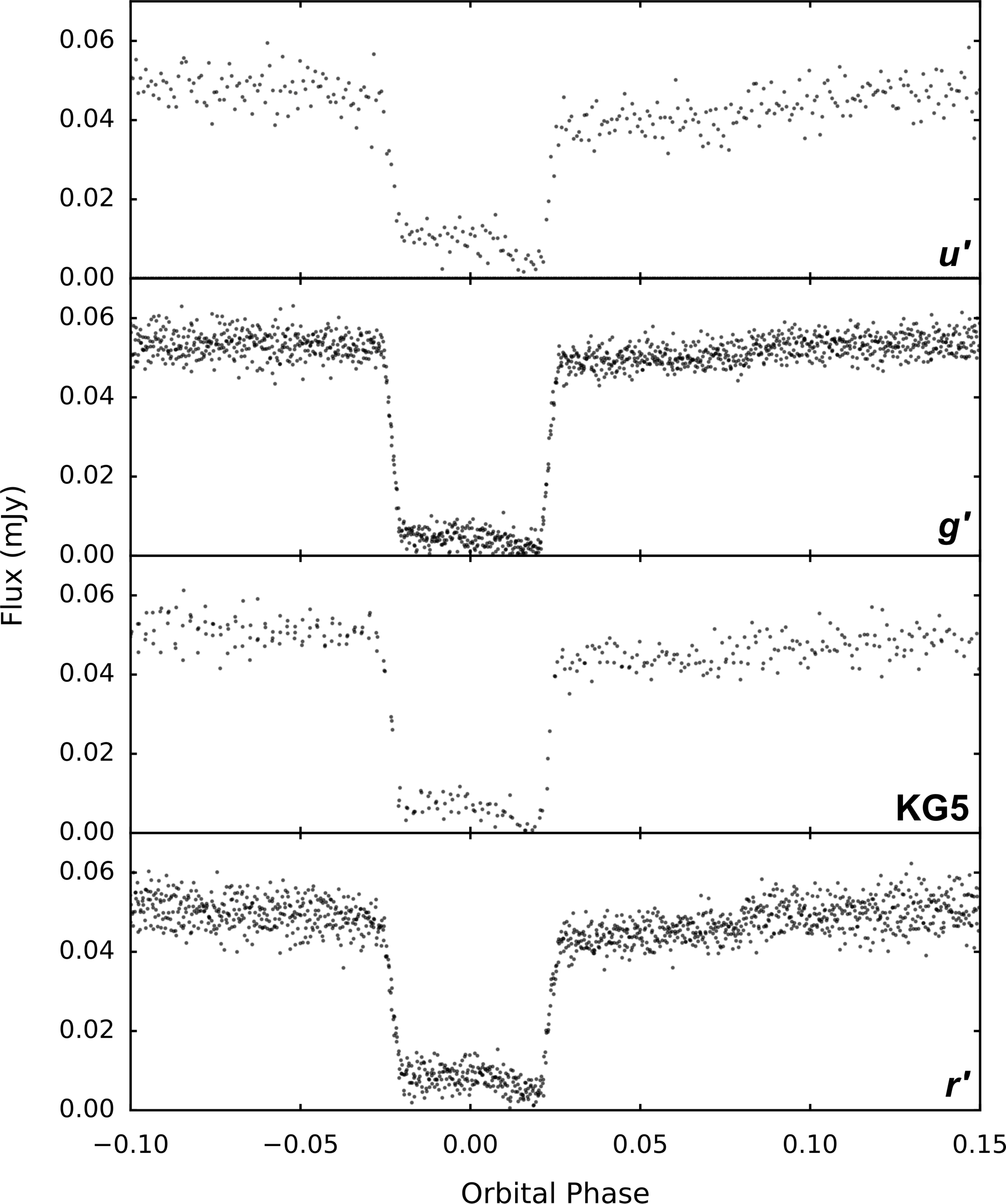}
\caption{\label{fig:ecl} Selected SDSS 1057 eclipses (phase-folded and overlaid) used to create average eclipses in each of the four wavelength bands. The name of each band is shown in the bottom-right corner of each plot.}
\end{center}
\end{figure}

\subsection{Light curve morphology and variations}
\label{subsec:lcmorph} 

All observations listed in Table~\ref{table:obs} show a clear white dwarf eclipse, while only a select few show a very faint bright spot eclipse. The difficulty in locating the bright spot eclipse feature in these light curves is due to the bright spot in SDSS 1057 being significantly less luminous than the white dwarf. This is made even harder due to the low signal-to-noise of each light curve -- a consequence of SDSS 1057 being a faint system ($g'\sim19.5$). In order to increase the signal-to-noise and strengthen the bright spot eclipse features, multiple eclipses have to be averaged. As discussed in \cite{McAllister17}, eclipse averaging can lead to inaccuracies if there are significant changes in disc radius. Such changes can shift the timing of the bright spot eclipse features over time and result in the broadening and weakening of these features after eclipse averaging. Not all systems exhibit significant disc radius changes, and visual analysis of the positions of the bright spot in individual eclipses show SDSS 1057 to have a constant disc radius -- making eclipse averaging suitable in this case.

The eclipses selected to contribute to the average eclipse in each wavelength band are phase-folded and plotted on top of each other in Figure~\ref{fig:ecl}. These include four out of the six ULTRACAM $u' g' r'$ eclipses and three out of the six ULTRASPEC KG5 eclipses. The 30 Dec 2013 and 23 Jun 2015 ULTRACAM observations were not included due to being affected by transparency variations, while the first three ULTRASPEC observations were not used due to a low signal-to-noise caused by overly short exposure times. As can be seen in Figure~\ref{fig:ecl}, there is no obvious flickering component in any SDSS 1057 eclipse light curve, but a large amount of white noise. Despite this, there are hints of a bright spot ingress feature around phase 0.01 and an egress at approximately phase 0.08. These features are clearest in the $r'$ band.

The resulting average eclipses in each band are shown in Figure~\ref{fig:all_fits}. All four eclipse light curves have seen an increase in signal-to-noise through averaging, and as a result the bright spot features have become clearer -- sufficiently so for eclipse model fitting (see section~\ref{subsec:avlcmod}). The sharp bright spot egress feature in the $r'$ band eclipse is further evidence for no significant disc radius changes in SDSS 1057 and validates the use of eclipse averaging in this instance.

\subsection{Simultaneous average light curve modelling}
\label{subsec:avlcmod}

\begin{figure*}
\begin{center}
\includegraphics[width=2.0\columnwidth,trim=100 30 100 50]{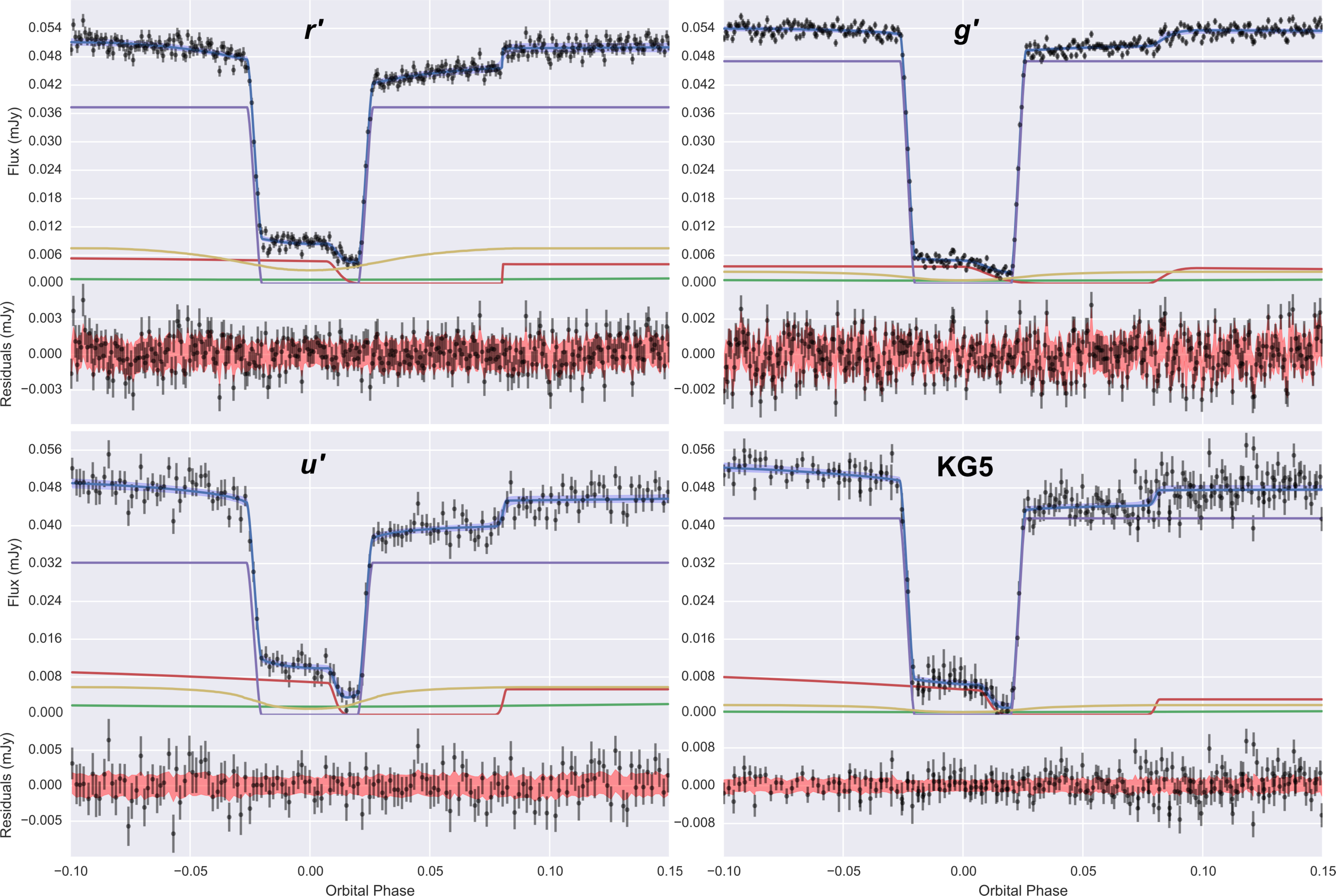}
\caption{\label{fig:all_fits} Simultaneous model fit (blue) to four averaged SDSS 1057 eclipses (black). The blue fill-between region represents 1$\sigma$ from the posterior mean of a random sample (size 1000) of the MCMC chain. Also shown are the different components of the model: white dwarf (purple), bright spot (red), accretion disc (yellow) and donor (green). The residuals are shown at the bottom of each plot, with the red fill-between region covering 2$\sigma$ from the posterior mean of the GP.}
\end{center}
\end{figure*}

The model of the binary system used to calculate eclipse light curves contains contributions from the white dwarf, bright spot, accretion disc and donor star, and is described in detail by \cite{savoury11}. The model makes a number of important assumptions: the bright spot lies on a ballistic trajectory from the donor, the donor fills its Roche lobe, the white dwarf is accurately described by a theoretical mass-radius relation, and an unobscured white dwarf \citep{savoury11}. The validity of this final assumption has been questioned by \cite{sparkodonoghue15} through fast photometry observations of the dwarf nova OY Car. However, as stated in \cite{mcallister15}, we feel this is still a reasonable assumption to make due to agreement between photometric and spectroscopic parameter estimates \citep{copperwheat12,savoury12}. Due to the tenuous bright spot in SDSS 1057, a simple bright spot model was preferred in this instance, with the four additional complex bright spot parameters introduced by \cite{savoury11} not included. The simple bright spot model was also chosen for modelling the eclipsing CV PHL 1445, another system with a weak bright spot \citep{mcallister15}.

As outlined in \cite{McAllister17}, our eclipse model has recently received two major modifications. First, it is now possible to fit multiple eclipse light curves simultaneously, whilst sharing parameters intrinsic to the system being modelled, e.g. mass ratio ($q$), white dwarf eclipse phase full-width at half-depth ($\Delta\phi$) and white dwarf radius ($R_{w}$) between all eclipses. Second, there is the option for any flickering present in the eclipse light curves to now also be modelled, thanks to the inclusion of an additional Gaussian process (GP) component. This requires three further parameters to the model, which represent the hyperparameters of the GP. For more details about the implementation of this additional GP component to the model, see \cite{McAllister17}. While the SDSS 1057 average light curves do not show any obvious signs of flickering, there is evidence for slight correlation in the residuals and therefore GPs are included in the analysis.

\begin{figure*}
\begin{center}
\includegraphics[width=1.8\columnwidth,trim=20 10 20 20]{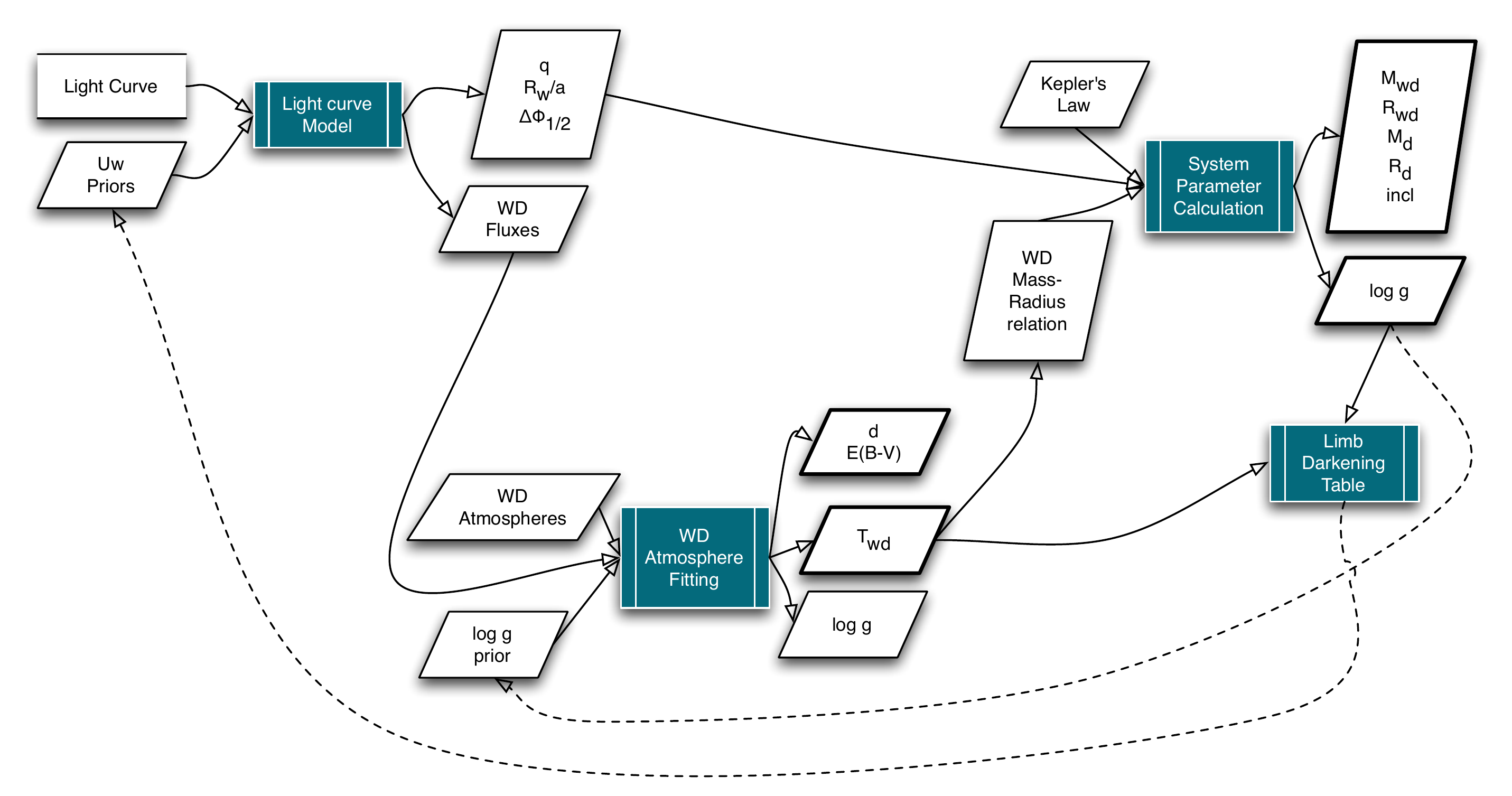}
\caption{\label{fig:flowchart} A schematic of the eclipse fitting procedure used to obtain system parameters. Two iterations of the fitting procedure occur, the dotted lines show steps to be taken only during the first iteration.}
\end{center}
\end{figure*}

The four average SDSS 1057 eclipses were fit simultaneously with the model -- GP component included. All 50 parameters were left to fit freely, except for the four limb-darkening parameters ($U_{\mathrm{w}}$). This is due to the data not being of sufficient quality to constrain values of $U_{\mathrm{w}}$ accurately. The $U_{\mathrm{w}}$ parameters' priors were heavily constrained around values inferred from the white dwarf temperature and log\,$g$ (see end of section~\ref{subsubsec:wdatmos}). These white dwarf parameters were determined through a preliminary run of the fitting procedure described throughout this section and shown schematically in Figure~\ref{fig:flowchart}.

An affine-invariant MCMC ensemble sampler \citep{goodmanweare10,foreman-mackey13} was used to draw samples from the posterior probability distribution of the model parameters. The MCMC was run for a total of 30,000 steps, with the first 20,000 of these used as part of a burn-in phase and discarded. The model fit to all four average eclipses is shown in Figure~\ref{fig:all_fits}. The blue line represents the most probable fit, and has a $\chi^{2}$ of 1561 with 966 degrees of freedom. The lines below each eclipse represent the separate components to the model: white dwarf (purple), bright spot (red), accretion disc (yellow) and donor (green). In addition to the most probable fit, a blue fill-between region can also be seen plotted on each eclipse. This represents 1$\sigma$ from the posterior mean of a random sample (size 1000) of the MCMC chain.

In all four eclipses, the model manages to fit both the white dwarf and bright spot eclipses successfully. There is no structure visible in the residuals at the phases corresponding to any of the ingresses and egresses.  In general, there is some structure in the residuals, which validates our decision to include the GP component. This component can be visualised through the red fill-between regions overlaying each eclipse's residuals in Figure~\ref{fig:all_fits}, and represents 2$\sigma$ from the GP's posterior mean. The GPs appear to model the residuals successfully in the $r'$ and $g'$ bands, but struggles for $u'$ and KG5. This may be due to differing amplitudes and timescales of the noise between eclipses, while our GP component can currently only accommodate for a shared amplitude and timescale between all eclipses.

\subsubsection{White dwarf atmosphere fitting}
\label{subsubsec:wdatmos}

The depths of the four white dwarf eclipses from the simultaneous fit provide a measure of the white dwarf flux at $u'$, $g'$, $r'$ and KG5 wavelengths. Estimates of the white dwarf temperature, log\,$g$ and distance were obtained through fitting these white dwarf fluxes to white dwarf atmosphere predictions \citep{bergeron95} with an affine-invariant MCMC ensemble sampler \citep{goodmanweare10,foreman-mackey13}. Reddening was also included as a parameter, in order for its uncertainty to be taken into account, but is not constrained by our data. Its prior covered the range from 0 to the maximum galactic extinction along the line-of-sight \citep{schlaflyfinkbeiner11}. The white dwarf fluxes and errors were taken as median values and standard deviations from a random sample of the simultaneous eclipse fit chain. A 3\% systematic error was added to the fluxes to account for uncertainties in photometric calibration.

Knowledge of the white dwarf temperature and log\,$g$ values enabled the estimation of the $U_{\mathrm{w}}$ parameters, with use of the data tables in \cite{gianninas13}. Linear limb-darkening parameters of 0.427, 0.392 and 0.328 were determined for the $u'$, $g'$ and $r'$ bands, respectively. A value of 0.374 for the KG5 band was calculated by taking a weighted mean of the $u'$, $g'$ and $r'$ values, based on the approximate fraction of the KG5 bandpass covered by each of the three SDSS filters.

\subsubsection{System parameters}
\label{subsubsec:sysparams}

The posterior probability distributions of $q$, $\Delta\phi$ and $R_{\mathrm{w}}/a$ returned by the MCMC eclipse fit described in section~\ref{subsec:avlcmod} were used along with Kepler's third law, the system's orbital period and a temperature-corrected white dwarf mass-radius relationship \citep{wood95}, to calculate the posterior probability distributions of the system parameters \citep{savoury11}, which include:

\begin{enumerate}
\item mass ratio, $q$;
\item white dwarf mass, $M_{\mathrm{w}}$;
\item white dwarf radius, $R_{\mathrm{w}}$;
\item white dwarf log\,$g$;
\item donor mass, $M_{\mathrm{d}}$;
\item donor radius, $R_{\mathrm{d}}$;
\item binary separation, $a$;
\item white dwarf radial velocity, $K_{\mathrm{w}}$;
\item donor radial velocity, $K_{\mathrm{d}}$;
\item inclination, $i$.
\end{enumerate}

The most likely value of each distribution is taken as the value of each system parameter, with upper and lower bounds derived from 67\% confidence levels.

\begin{figure}
\begin{center}
\includegraphics[width=1.0\columnwidth,trim=20 10 20 10]{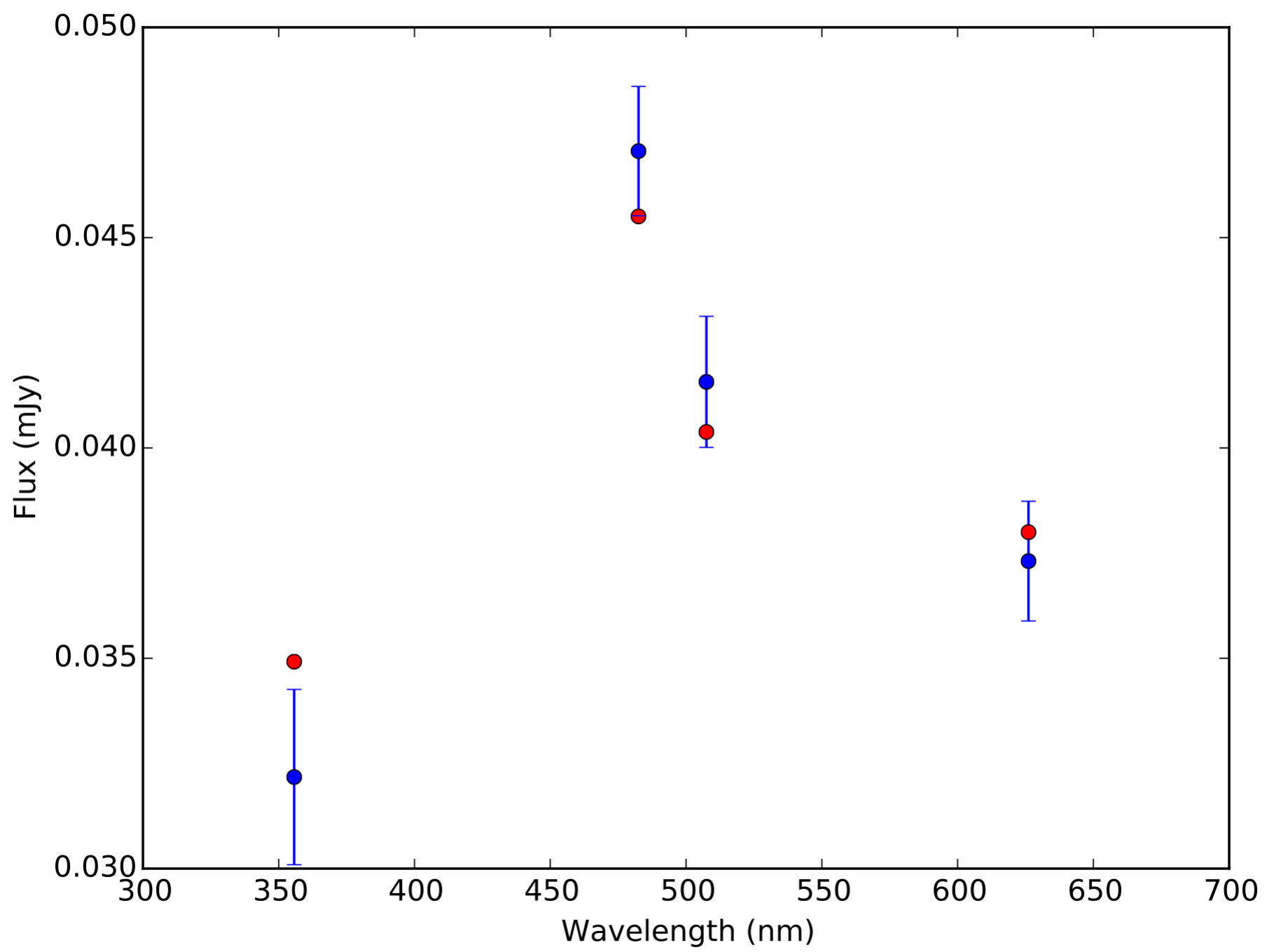}
\caption{\label{fig:fluxes} White dwarf fluxes from the simultaneous 4-eclipse model fit (blue) and \protect\cite{bergeron95} white dwarf atmosphere predictions (red), at wavelengths corresponding to (from left to right) $u'$, $g'$, KG5 and $r'$ filters.}
\end{center}
\end{figure}

There are two iterations to the fitting procedure (Figure~\ref{fig:flowchart}), with system parameters calculated twice in total. The value for log\,$g$ returned from the first calculation was used to constrain the log\,$g$ prior in a second MCMC fit of the model atmosphere predictions \citep{bergeron95} to the white dwarf fluxes, as described in section~\ref{subsubsec:wdatmos}. The results of this MCMC fit can be found in Figure~\ref{fig:fluxes}, with the measured white dwarf fluxes in each band in blue and the white dwarf atmosphere model in red. The model and fluxes are in good agreement in all wavelength bands, however it appears that the measured $u'$ band flux is slightly underestimated. On close inspection of the $u'$ band eclipse fit in Figure~\ref{fig:all_fits}, we find a greater than expected contribution from both the disc and donor at this wavelength, opening up the possibility that a small fraction of the true white dwarf flux may have been mistakenly attributed to these components. The measured fluxes from SDSS 1057 are consistent with a white dwarf of temperature 13300\,$\pm$\,1100\,K and distance 367\,$\pm$\,26\,pc.

\begin{figure}
\begin{center}
\includegraphics[width=1.0\columnwidth,trim=20 10 20 10]{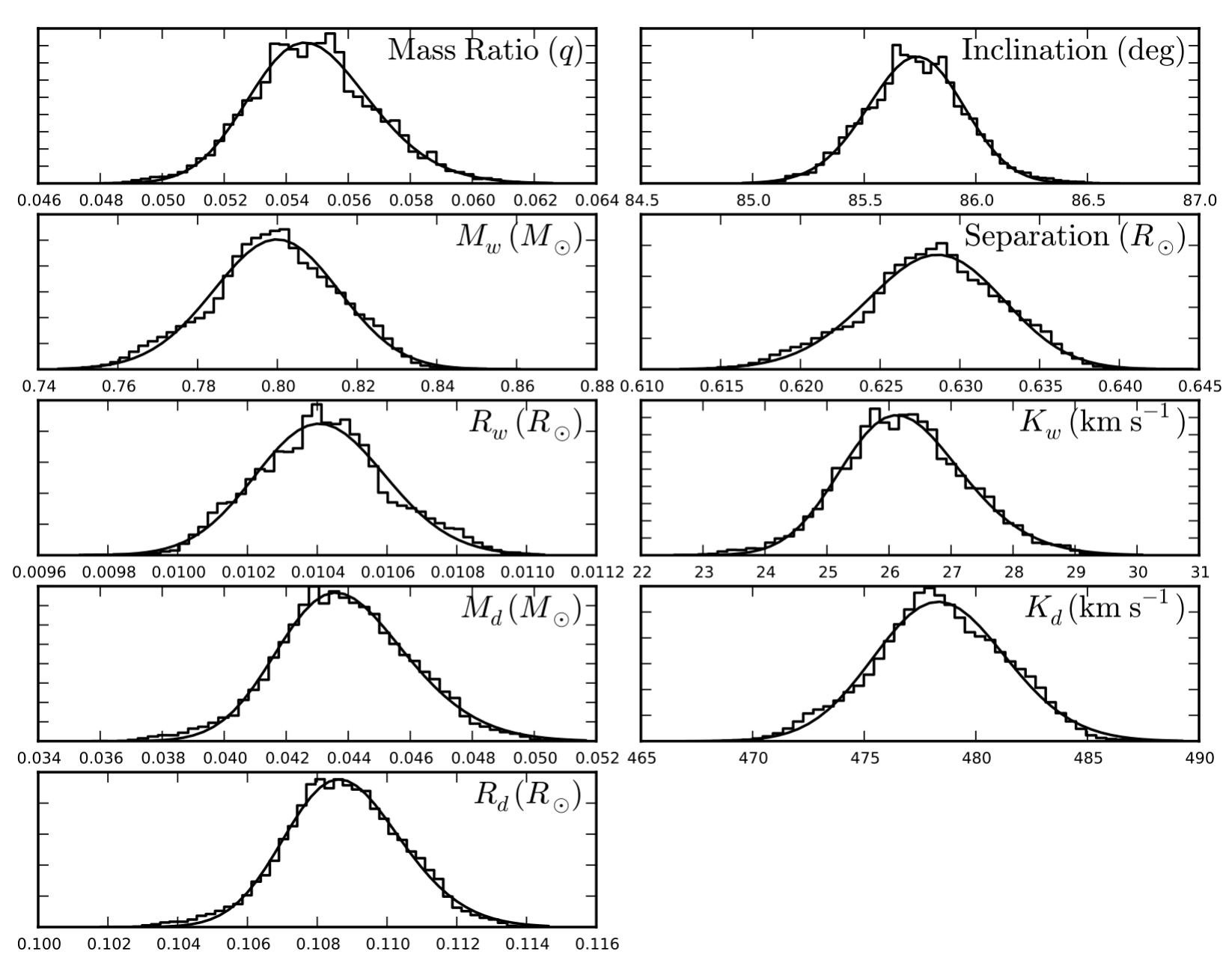}
\caption{\label{fig:pdf} Normalised posterior probability density function for each system parameter.}
\end{center}
\end{figure}

The posterior probability distributions of the system parameters are shown in Figure~\ref{fig:pdf}, while their calculated values are given in Table~\ref{table:syspars}. Also included in Table~\ref{table:syspars} are the estimates of the white dwarf temperature and distance from the white dwarf atmosphere fitting.

\begingroup
\setlength{\tabcolsep}{10pt}
\renewcommand{\arraystretch}{1.5}
\begin{table}
\begin{center}
\begin{tabular}{lc}
\hline
\hline
$q$ & 0.0546\,$\pm$\,0.0020 \\
$M_{w}$ (M$_{\odot}$) & 0.800\,$\pm$\,0.015 \\
$R_{w}$ (R$_{\odot}$) & 0.01040\,$\pm$\,0.00017 \\
$M_{d}$ (M$_{\odot}$) & 0.0436\,$\pm$\,0.0020 \\
$R_{d}$ (R$_{\odot}$) & 0.1086\,$\pm$\,0.0017 \\
$a$ (R$_{\odot}$) & 0.629\,$\pm$\,0.004 \\
$K_{w}$ (km\,s$^{-1}$) & 26.2\,$^{+1.1}_{-0.8}$ \\
$K_{d}$ (km\,s$^{-1}$) & 478\,$\pm$\,3 \\
$i$ $(^{\circ})$ & 85.74\,$\pm$\,0.21 \\
log\,$g$ & 8.307\,$\pm$\,0.017 \\ \hline
$T_{w}$ (K) & 13300\,$\pm$\,1100 \\
$d$ (pc) & 367\,$\pm$\,26 \\
\hline
\hline
\end{tabular}
\caption{\label{table:syspars} System parameters for SDSS 1057. $T_{\mathrm{w}}$ and $d$ represent the temperature and distance of the white dwarf, respectively.}
\end{center}
\end{table}
\endgroup

\subsubsection{Spectral energy distribution}
\label{subsubsec:sed}

\cite{southworth15} use both the SDSS spectrum and GALEX fluxes \citep{morrissey07} to analyse the spectral energy distribution of SDSS 1507. The model of \cite{gaensicke06} is able to successfully reproduce the SDSS spectrum with a white dwarf temperature of 10500\,K, log\,$g$ of 8.0, distance of 305\,pc, accretion disc temperature of 5800\,K and an L5 secondary star. However, the model does not provide a good fit to the GALEX fluxes, which \cite{southworth15} state could have been taken during eclipse.

\begin{figure}
\begin{center}
\includegraphics[width=1.0\columnwidth,trim=20 10 20 10]{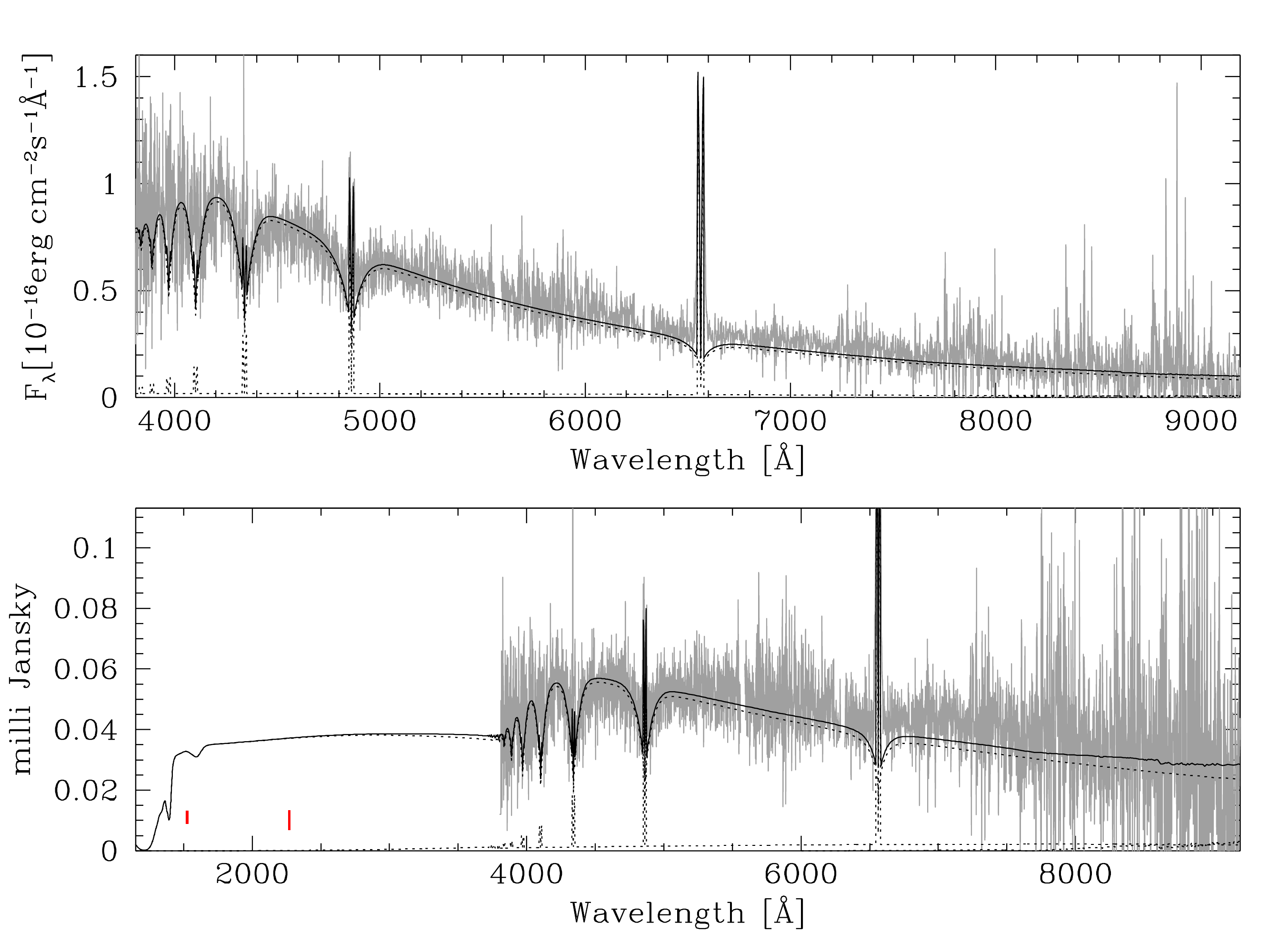}
\caption{\label{fig:sdss} Three-component model spectra (black) overlaid on top of the SDSS spectrum of SDSS 1057 (grey). The three components include a white dwarf, an isothermal and isobaric hydrogen slab and a mid-T secondary star. The two red data points represent UV flux measurements from GALEX.}
\end{center}
\end{figure}

As we arrive at a slightly different white dwarf temperature, log\,$g$ and distance (Table~\ref{table:syspars}), as well as a slightly later spectral type secondary, we investigated whether the \cite{gaensicke06} model with these parameters is still a good fit to the SDSS spectrum. The resulting fit is shown in Figure~\ref{fig:sdss}. While the fit is good, the white dwarf temperature used appears to produce a slope that is slightly too blue, hinting that it might be marginally overestimated, but this may be corrected with alternate disc parameters. As in \cite{southworth15}, the GALEX fluxes (red data points) are again not fit well by the model, with both the near- and far-UV fluxes much lower than predicted. Using the ephemeris in Equation~\ref{eq:ephem}, we can rule out the possibility of these fluxes being taken during eclipse. Another reason for these low UV flux measurements could be due to absorption by an ``accretion veil" of hot gas positioned above the accretion disc \citep{horne94,copperwheat12}. This explanation consequently invalidates our prior assumption of an unobscured white dwarf (see Section~\ref{subsec:avlcmod}). However, we can take reassurance from the agreement between photometric and spectroscopic parameter estimates for two eclipsing CVs (OY Car and CTCV J1300-3052) that both show convincing evidence for an accretion veil \citep{copperwheat12,savoury12}.

\section{Discussion}
\label{sec:discussion}

\subsection{Component masses}
\label{subsec:masses}

The white dwarf in SDSS 1057 is found to have a mass of 0.800\,$\pm$\,0.015\,M$_{\odot}$, which is close to the mean CV white dwarf mass of 0.81\,$\pm$\,0.04\,M$_{\odot}$ \citep{savoury11} but notably higher than both the mean post-common-envelope binary (PCEB) white dwarf mass of 0.58\,$\pm$\,0.20\,M$_{\odot}$\citep{zorotovic11} and mean white dwarf field mass of 0.621\,M$_{\odot}$ \citep{tremblay16}.

The donor has a mass of $0.0436\,\pm\,0.0020\,$M$_{\odot}$, which makes it not only substellar -- as it is well below the hydrogen burning limit of $\sim0.075\,$M$_{\odot}$ \citep{kumar63,hayashinakano63} -- but also the lowest mass donor yet measured in an eclipsing CV.

\subsection{Mass transfer rate}
\label{subsec:mtr}

We calculate a medium-term average mass transfer rate of $\dot{M}\,=\,6.0\,^{+2.9}_{-2.1}\,\times\,10^{-11}\,$M$_{\odot}\,$yr$^{-1}$ using the white dwarf mass and temperature \citep{townsleybildsten03,townsleygansicke09}. This is a number of times greater than the expected secular mass transfer rate of $\dot{M}\,\sim\,1.5\,\times\,10^{-11}\,$M$_{\odot}\,$yr$^{-1}$ for a period-bounce system at this orbital period \citep{knigge11}, and is actually consistent with the secular mass transfer rate of a \textit{pre}-bounce system of the same orbital period. This is further evidence that the white dwarf temperature we derive through white dwarf atmosphere predictions may be slightly overestimated.

Recalculating the medium-term average mass transfer rate using the lower white dwarf temperature of 10500\,K from \cite{southworth15} brings it much more in line with the expected secular mass transfer rate. Importantly, the system parameters we obtain are consistent within errors, regardless of whether a white dwarf temperature of 10500\,K or 13300\,K is used to correct the white dwarf mass-radius relationship.

\subsection{White dwarf pulsations}
\label{subsec:pulse}

The white dwarf's temperature and log\,$g$ put it just outside the blue edge of the DAV instability strip, which opens up the possibility of pulsations \citep{gianninas11}. The lack of out-of-eclipse coverage and low signal-to-noise of this data is not conducive to a search for pulsations, and therefore out-of-eclipse follow-up observations are required to determine whether this white dwarf is pulsating.

\subsection{Evolutionary state of SDSS 1057}
\label{subsec:evostate}

\begin{figure*}
\begin{center}
\includegraphics[width=1.6\columnwidth,trim=80 30 80 30]{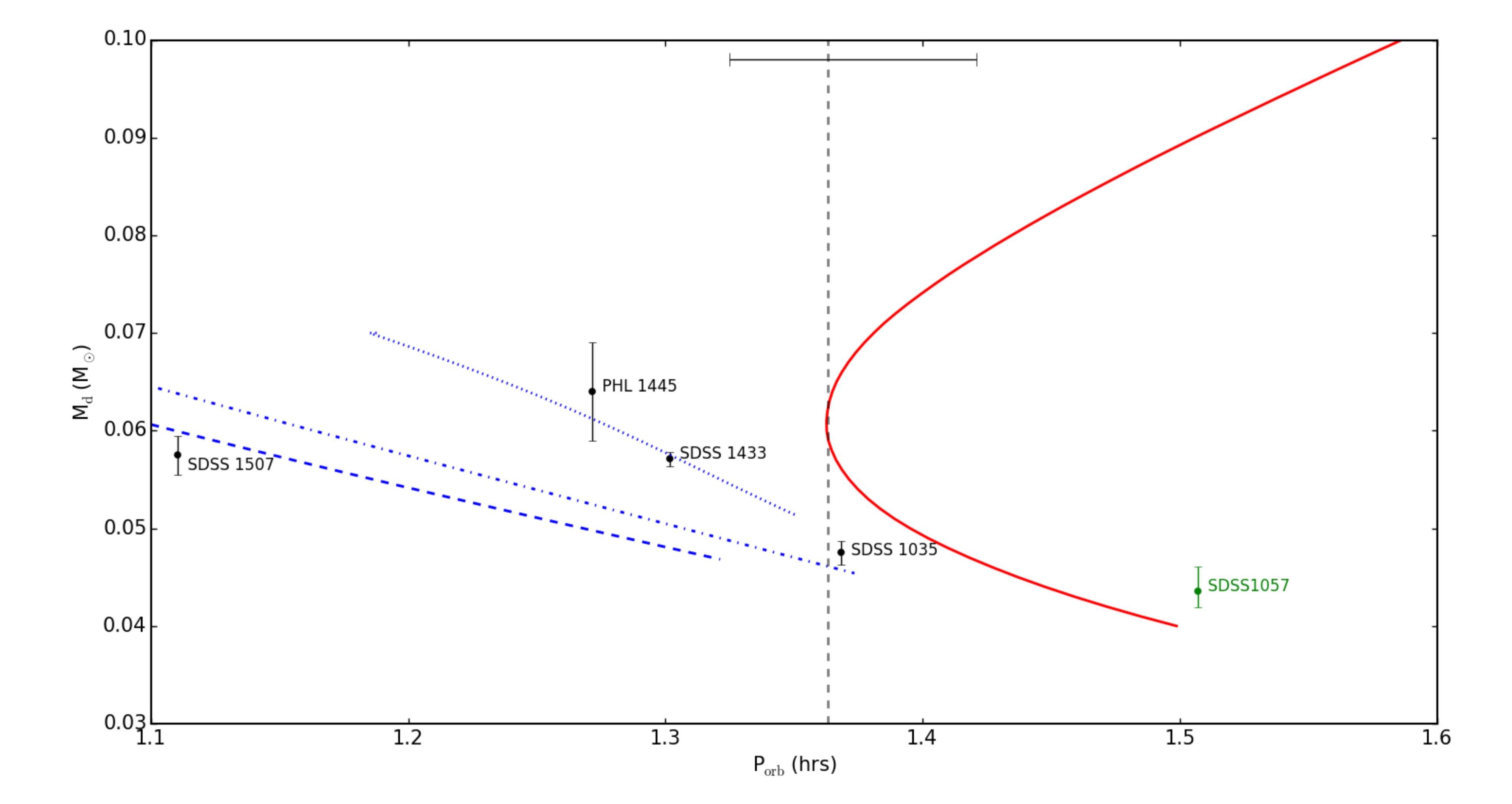}
\caption{\label{fig:m2vsP}Donor mass ($M_{\mathrm{d}}$) vs orbital period ($P_{\mathrm{orb}}$) for SDSS 1057 and other substellar donor eclipsing CVs: SDSS 1433, SDSS 1035, SDSS 1507 and PHL 1445 \citep{savoury11,mcallister15}. Also plotted are evolutionary tracks for both main-sequence (red line; \protect\citealt{knigge11}) and brown dwarf (blue lines; \protect\citealt{mcallister15}) donors. The three brown dwarf donor tracks vary in donor age at start of mass transfer, with the dashed, dot-dashed and dotted lines representing 2\,Gyr, 1\,Gyr and 600\,Myr, respectively. The vertical dashed line represents the location of the CV period minimum determined by \protect\cite{knigge11}, with the shaded area representing the error on this value. The bar across the top of the plot shows the FWHM of the CV period spike observed by \protect\cite{gaensicke09}.}
\end{center}
\end{figure*}

The relation between donor mass and orbital period in CVs was used to investigate the evolutionary status of SDSS 1057. Figure~\ref{fig:m2vsP} shows SDSS 1057's donor mass ($M_{\mathrm{d}}$) plotted against orbital period ($P_{\mathrm{orb}}$), along with the four other known substellar donor eclipsing systems: SDSS J150722.30+523039.8 (SDSS 1507), PHL 1445, SDSS J143317.78+101123.3 (SDSS 1433) and SDSS J103533.03+055158.4 (SDSS 1035) \citep{savoury11,mcallister15}. Also plotted are four evolutionary tracks: a red track representing the evolution of a CV with a main-sequence donor \citep{knigge11}, and three blue tracks as examples of evolution when systems contain a brown dwarf donor from formation \citep{mcallister15}.

CV systems that follow the main-sequence track evolve from longer to shorter periods -- right to left in Figure~\ref{fig:m2vsP} -- until the orbital period minimum (vertical dashed line) is reached, at which point they head back towards longer periods. Systems that form with a brown dwarf donor instead start at shorter periods and evolve to longer periods -- left to right in Figure~\ref{fig:m2vsP} -- and eventually join up with the post-period-bounce main-sequence track. The three brown dwarf donor tracks shown in Figure~\ref{fig:m2vsP} all have the same initial white dwarf (0.75\,M$_{\odot}$) and donor (0.07\,M$_{\odot}$) masses, but have different donor ages at start of mass transfer. The dashed, dot-dashed and dotted blue lines represent donor ages of 2\,Gyr, 1\,Gyr and 600\,Myr respectively.

Figure~\ref{fig:m2vsP} is similar to Figure 9 from \cite{mcallister15}, but now with SDSS 1057 added in. The evolutionary status of each of the four existing substellar systems were discussed in detail in \cite{mcallister15}, which we summarise here. SDSS 1507 lies significantly below the period minimum in Figure~\ref{fig:m2vsP} due to being metal poor as a member of the Galactic halo, inferred from SDSS 1507's high proper motion \citep{patterson08,uthas11}. This is an exceptional system and therefore we do not include it in the remaining discussion. From their positions in Figure~\ref{fig:m2vsP}, the best apparent explanation for PHL 1445 and SDSS 1433 (and arguably also SDSS 1035) is formation with a brown dwarf donor. However, due to the observation of a ``brown dwarf desert'' \citep{duquennoy91,marcy00,grether06} the progenitors of such systems -- and therefore the systems themselves -- are expected to be very rare and greatly outnumbered by those following the main-sequence track. This makes it unlikely for even a single one of these systems to have formed with a brown dwarf donor, never mind the majority of this (albeit small) sample. The most likely scenario is that all three systems belong to the main-sequence track, which raises concerns for the accuracy of this track (see Section~\ref{subsec:pmin}).

As it has the lowest donor mass of all other systems discussed above and an orbital period significantly greater than the period minimum, we find SDSS 1057 to be positioned close to the period-bounce arm of the main-sequence donor track in Figure~\ref{fig:m2vsP}. Its 90.44\,min period puts distance between itself and the period minimum, giving SDSS 1057 the best case for being a true period bouncer among the other currently known substellar systems. This is backed up by SDSS 1057 possessing additional period-bouncer traits: low white dwarf temperature (although at 13300\,K it is at the upper end of what's expected; \citealt{patterson11}), faint quiescent magnitude ($g'\simeq19.5$ at $d\simeq367$\,pc) and long outburst recurrence time (no outburst recorded in over 8 years of CRTS observations; \citealt{drake09}). It must be stated that due to the merging of the brown dwarf and main-sequence donor tracks post-period minimum, the scenario of SDSS 1057 directly forming with a brown dwarf donor cannot be  ruled out. However, due to the lack of potential progenitors and with 80\% predicted to lie below the period minimum \citep{politano04}, this seems unlikely to be the case.

\subsection{CV Evolution at period minimum}
\label{subsec:pmin}

This study of SDSS 1057 brings the total number of modelled eclipsing period-minimum/period-bounce systems -- and therefore systems with precise system parameters -- to seven. This includes the period minimum systems SDSS J150137.22+550123.3 (SDSS 1501), SDSS J090350.73+330036.1 (SDSS 0903) and SDSS J150240.98+333423.9 (SDSS 1502) from \cite{savoury11}, which all have periods \textless\,86\,min but aren't included in Figure~\ref{fig:m2vsP} due to having donor masses above the substellar limit.

It is evident that none of these systems -- including SDSS 1057 -- lie on the main-sequence donor track itself, with some (namely PHL 1445 and SDSS 1433) located far from it. This raises questions about the accuracy of the donor track in the period minimum regime, but it may be the case that there is a large intrinsic scatter associated with the track. It is expected for a small amount of intrinsic scatter to exist due to differences in white dwarf mass, but a significant contribution may come from variations in the additional angular momentum loss (approximately 2.5$\times$ gravitational radiation) that is required in order for the donor track to conform with the observed period minimum \citep{gaensicke09,knigge11}. In \cite{mcallister15} we used the width of the observed period minimum from \cite{gaensicke09} as a measure of the intrinsic scatter of the main-sequence donor track, but we concluded this was too small to account for the positions of PHL 1445 and SDSS 1433.

With such a small sample of observations currently available, it is not possible to thoroughly test the validity of the main-sequence donor evolutionary track at the period minimum. Many more precise masses from period-minimum/period-bounce systems are required, and therefore every additional eclipsing system within this regime that is suitable for modelling is of great value.

\section{Conclusions}
\label{sec:conclusions}

We have presented high-speed photometry of the faint eclipsing CV SDSS 1057. By increasing signal-to-noise through averaging multiple eclipses, a faint bright spot eclipse feature emerged from the white dwarf-dominated eclipse profiles. The presence of bright spot eclipse features enabled the determination of system parameters through fitting an eclipse model to average eclipses in four different wavelength bands simultaneously. Multi-wavelength observations allowed a white dwarf temperature and distance to be estimated through fits of model atmosphere predictions to white dwarf fluxes.

While the white dwarf in SDSS 1057 has a mass comparable to the average for CV white dwarfs, we find the donor to have the lowest mass of any known eclipsing CV donor. A low donor mass -- coupled with an orbital period significantly greater than the period minimum -- is strong evidence for SDSS 1057 being a bona fide period-bounce system, although formation from a white dwarf/brown dwarf binary cannot be ruled out. Every eclipsing period-minimum/period-bounce CV is of great interest, with so few systems with precise system parameters currently known. As a consequence, the evolution of systems in this regime is not yet fully understood.

\section{Acknowledgements}
\label{sec:acknowledgements}

We thank the anonymous referee for their comments. MJM acknowledges the support of a UK Science and Technology Facilities Council (STFC) funded PhD. SPL and VSD are supported by STFC grant ST/J001589/1. TRM and EB are supported by the STFC 
grant ST/L000733/1. VSD and TRM acknowledge the support of the Leverhulme Trust for the operation of ULTRASPEC at the Thai National Telescope. Support for this work was provided by NASA through Hubble Fellowship grant \#HST-HF2-51357.001-A. The research leading to these results has received funding from the
European Research Council under the European Union's Seventh Framework Programme (FP/2007-2013) / ERC Grant Agreement n. 320964 (WDTracer). The results presented are based on observations made with the William Herschel Telescope, operated at the Spanish Observatorio del Roque de Los Muchachos of the Instituto de Astrof\'isica de Canarias by the Isaac Newton Group, as well as the Thai National Telescope, operated at the Thai National Observatory by the National Astronomical Research Institute of Thailand. This research has made use of NASA's Astrophysics Data System Bibliographic Services.

\bibliography{sdss1057_references}

\begin{thebibliography}{48}
\expandafter\ifx\csname natexlab\endcsname\relax\def\natexlab#1{#1}\fi

\bibitem[{{Bell} {et~al}\mbox{.}(2012){Bell}, {Naylor}, {Mayne}, {Jeffries}, \&
  {Littlefair}}]{bell12}
{Bell} C.~P.~M., {Naylor} T., {Mayne} N.~J., {Jeffries} R.~D., {Littlefair}
  S.~P., 2012, \mnras, 424, 3178

\bibitem[{{Bergeron}, {Wesemael} \& {Beauchamp}(1995){Bergeron}, {Wesemael}, \&
  {Beauchamp}}]{bergeron95}
{Bergeron} P., {Wesemael} F., {Beauchamp} A., 1995, \pasp, 107, 1047

\bibitem[{{Copperwheat} {et~al}\mbox{.}(2012){Copperwheat}, {Marsh}, {Parsons},
  {Hickman}, {Steeghs}, {Breedt}, {Dhillon}, {Littlefair}, \&
  {Savoury}}]{copperwheat12}
{Copperwheat} C.~M. {et~al.}, 2012, \mnras, 421, 149

\bibitem[{{Dhillon} {et~al}\mbox{.}(2014){Dhillon}, {Marsh}, {Atkinson},
  {Bezawada}, {Bours}, {Copperwheat}, {Gamble}, {Hardy}, {Hickman}, {Irawati},
  {Ives}, {Kerry}, {Leckngam}, {Littlefair}, {McLay}, {O'Brien}, {Peacocke},
  {Poshyachinda}, {Richichi}, {Soonthornthum}, \& {Vick}}]{dhillon14}
{Dhillon} V.~S. {et~al.}, 2014, \mnras, 444, 4009

\bibitem[{{Dhillon} {et~al}\mbox{.}(2007){Dhillon}, {Marsh}, {Stevenson},
  {Atkinson}, {Kerry}, {Peacocke}, {Vick}, {Beard}, {Ives}, {Lunney}, {McLay},
  {Tierney}, {Kelly}, {Littlefair}, {Nicholson}, {Pashley}, {Harlaftis}, \&
  {O'Brien}}]{dhillon07}
{Dhillon} V.~S. {et~al.}, 2007, \mnras, 378, 825

\bibitem[{{Drake} {et~al}\mbox{.}(2009){Drake}, {Djorgovski}, {Mahabal},
  {Beshore}, {Larson}, {Graham}, {Williams}, {Christensen}, {Catelan},
  {Boattini}, {Gibbs}, {Hill}, \& {Kowalski}}]{drake09}
{Drake} A.~J. {et~al.}, 2009, \apj, 696, 870

\bibitem[{{Duquennoy} \& {Mayor}(1991)}]{duquennoy91}
{Duquennoy} A., {Mayor} M., 1991, \aap, 248, 485

\bibitem[{{Foreman-Mackey} {et~al}\mbox{.}(2013){Foreman-Mackey}, {Hogg},
  {Lang}, \& {Goodman}}]{foreman-mackey13}
{Foreman-Mackey} D., {Hogg} D.~W., {Lang} D., {Goodman} J., 2013, \pasp, 125,
  306

\bibitem[{{G{\"a}nsicke} {et~al}\mbox{.}(2009){G{\"a}nsicke}, {Dillon},
  {Southworth}, {Thorstensen}, {Rodr{\'{\i}}guez-Gil}, {Aungwerojwit}, {Marsh},
  {Szkody}, {Barros}, {Casares}, {de Martino}, {Groot}, {Hakala}, {Kolb},
  {Littlefair}, {Mart{\'{\i}}nez-Pais}, {Nelemans}, \&
  {Schreiber}}]{gaensicke09}
{G{\"a}nsicke} B.~T. {et~al.}, 2009, \mnras, 397, 2170

\bibitem[{{G{\"a}nsicke} {et~al}\mbox{.}(2006){G{\"a}nsicke},
  {Rodr{\'{\i}}guez-Gil}, {Marsh}, {de Martino}, {Nestoras}, {Szkody},
  {Aungwerojwit}, {Barros}, {Dillon}, {Araujo-Betancor}, {Ar{\'e}valo},
  {Casares}, {Groot}, {Kolb}, {L{\'a}zaro}, {Hakala}, {Mart{\'{\i}}nez-Pais},
  {Nelemans}, {Roelofs}, {Schreiber}, {van den Besselaar}, \&
  {Zurita}}]{gaensicke06}
{G{\"a}nsicke} B.~T. {et~al.}, 2006, \mnras, 365, 969

\bibitem[{{Gianninas}, {Bergeron} \& {Ruiz}(2011){Gianninas}, {Bergeron}, \&
  {Ruiz}}]{gianninas11}
{Gianninas} A., {Bergeron} P., {Ruiz} M.~T., 2011, \apj, 743, 138

\bibitem[{{Gianninas} {et~al}\mbox{.}(2013){Gianninas}, {Strickland}, {Kilic},
  \& {Bergeron}}]{gianninas13}
{Gianninas} A., {Strickland} B.~D., {Kilic} M., {Bergeron} P., 2013, \apj, 766,
  3

\bibitem[{{Goliasch} \& {Nelson}(2015)}]{goliasch15}
{Goliasch} J., {Nelson} L., 2015, \apj, 809, 80

\bibitem[{{Goodman} \& {Weare}(2010)}]{goodmanweare10}
{Goodman} J., {Weare} J., 2010, Comm. App. Math. Comp. Sci., 5, 65

\bibitem[{{Grether} \& {Lineweaver}(2006)}]{grether06}
{Grether} D., {Lineweaver} C.~H., 2006, \apj, 640, 1051

\bibitem[{{Hardy} {et~al}\mbox{.}(2016){Hardy}, {McAllister}, {Dhillon},
  {Littlefair}, {Bours}, {Breedt}, {Butterley}, {Chakpor}, {Irawati}, {Kerry},
  {Marsh}, {Parsons}, {Savoury}, {Wilson}, \& {Woudt}}]{hardy16}
{Hardy} L.~K. {et~al.}, 2016, ArXiv e-prints

\bibitem[{{Hayashi} \& {Nakano}(1963)}]{hayashinakano63}
{Hayashi} C., {Nakano} T., 1963, Progress of Theoretical Physics, 30, 460

\bibitem[{{Hellier}(2001)}]{hellier01}
{Hellier} C., 2001, {Cataclysmic Variable Stars: How and Why they Vary}.
  Springer-Praxis, New York

\bibitem[{{Horne} {et~al}\mbox{.}(1994){Horne}, {Marsh}, {Cheng}, {Hubeny}, \&
  {Lanz}}]{horne94}
{Horne} K., {Marsh} T.~R., {Cheng} F.~H., {Hubeny} I., {Lanz} T., 1994, \apj,
  426, 294

\bibitem[{{Knigge}, {Baraffe} \& {Patterson}(2011){Knigge}, {Baraffe}, \&
  {Patterson}}]{knigge11}
{Knigge} C., {Baraffe} I., {Patterson} J., 2011, \apjs, 194, 28

\bibitem[{{Kolb}(1993)}]{kolb93}
{Kolb} U., 1993, \aap, 271, 149

\bibitem[{{Kumar}(1963)}]{kumar63}
{Kumar} S.~S., 1963, \apj, 137, 1121

\bibitem[{{Littlefair} {et~al}\mbox{.}(2006){Littlefair}, {Dhillon}, {Marsh},
  {G{\"a}nsicke}, {Southworth}, \& {Watson}}]{littlefair06}
{Littlefair} S.~P., {Dhillon} V.~S., {Marsh} T.~R., {G{\"a}nsicke} B.~T.,
  {Southworth} J., {Watson} C.~A., 2006, Science, 314, 1578

\bibitem[{{Littlefair}, {Dhillon} \& {Mart{\'{\i}}n}(2003){Littlefair},
  {Dhillon}, \& {Mart{\'{\i}}n}}]{littlefair03}
{Littlefair} S.~P., {Dhillon} V.~S., {Mart{\'{\i}}n} E.~L., 2003, \mnras, 340,
  264

\bibitem[{{Marcy} \& {Butler}(2000)}]{marcy00}
{Marcy} G.~W., {Butler} R.~P., 2000, \pasp, 112, 137

\bibitem[{{McAllister} {et~al}\mbox{.}(2015){McAllister}, {Littlefair},
  {Baraffe}, {Dhillon}, {Marsh}, {Bento}, {Bochinski}, {Bours}, {Breedt},
  {Copperwheat}, {Hardy}, {Kerry}, {Parsons}, {Rostron}, {Sahman}, {Savoury},
  \& {Tunnicliffe}}]{mcallister15}
{McAllister} M.~J. {et~al.}, 2015, \mnras, 451, 114

\bibitem[{{McAllister} {et~al}\mbox{.}(2017){McAllister}, {Littlefair},
  {Dhillon}, {Marsh}, {Ashley}, {Bours}, {Breedt}, {Hardy}, {Hermes},
  {Kengkriangkrai}, {Kerry}, {Rattanasoon}, \& {Sahman}}]{McAllister17}
{McAllister} M.~J. {et~al.}, 2017, \mnras, 464, 1353

\bibitem[{{Morrissey} {et~al}\mbox{.}(2007){Morrissey}, {Conrow}, {Barlow},
  {Small}, {Seibert}, {Wyder}, {Budav{\'a}ri}, {Arnouts}, {Friedman},
  {Forster}, {Martin}, {Neff}, {Schiminovich}, {Bianchi}, {Donas}, {Heckman},
  {Lee}, {Madore}, {Milliard}, {Rich}, {Szalay}, {Welsh}, \&
  {Yi}}]{morrissey07}
{Morrissey} P. {et~al.}, 2007, \apjs, 173, 682

\bibitem[{{Patterson}(2011)}]{patterson11}
{Patterson} J., 2011, \mnras, 411, 2695

\bibitem[{{Patterson}, {Thorstensen} \& {Knigge}(2008){Patterson},
  {Thorstensen}, \& {Knigge}}]{patterson08}
{Patterson} J., {Thorstensen} J.~R., {Knigge} C., 2008, \pasp, 120, 510

\bibitem[{{Politano}(2004)}]{politano04}
{Politano} M., 2004, \apj, 604, 817

\bibitem[{{Rappaport}, {Joss} \& {Webbink}(1982){Rappaport}, {Joss}, \&
  {Webbink}}]{rappaport82}
{Rappaport} S., {Joss} P.~C., {Webbink} R.~F., 1982, \apj, 254, 616

\bibitem[{{Savoury} {et~al}\mbox{.}(2011){Savoury}, {Littlefair}, {Dhillon},
  {Marsh}, {G{\"a}nsicke}, {Copperwheat}, {Kerry}, {Hickman}, \&
  {Parsons}}]{savoury11}
{Savoury} C.~D.~J. {et~al.}, 2011, \mnras, 415, 2025

\bibitem[{{Savoury} {et~al}\mbox{.}(2012){Savoury}, {Littlefair}, {Marsh},
  {Dhillon}, {Parsons}, {Copperwheat}, \& {Steeghs}}]{savoury12}
{Savoury} C.~D.~J., {Littlefair} S.~P., {Marsh} T.~R., {Dhillon} V.~S.,
  {Parsons} S.~G., {Copperwheat} C.~M., {Steeghs} D., 2012, \mnras, 422, 469

\bibitem[{{Schlafly} \& {Finkbeiner}(2011)}]{schlaflyfinkbeiner11}
{Schlafly} E.~F., {Finkbeiner} D.~P., 2011, \apj, 737, 103

\bibitem[{{Smith} {et~al}\mbox{.}(2002){Smith}, {Tucker}, {Kent}, {Richmond},
  {Fukugita}, {Ichikawa}, {Ichikawa}, {Jorgensen}, {Uomoto}, {Gunn}, {Hamabe},
  {Watanabe}, {Tolea}, {Henden}, {Annis}, {Pier}, {McKay}, {Brinkmann}, {Chen},
  {Holtzman}, {Shimasaku}, \& {York}}]{smith02}
{Smith} J.~A. {et~al.}, 2002, \aj, 123, 2121

\bibitem[{{Southworth} {et~al}\mbox{.}(2015){Southworth}, {Tappert},
  {G{\"a}nsicke}, \& {Copperwheat}}]{southworth15}
{Southworth} J., {Tappert} C., {G{\"a}nsicke} B.~T., {Copperwheat} C.~M., 2015,
  \aap, 573, A61

\bibitem[{{Spark} \& {O'Donoghue}(2015)}]{sparkodonoghue15}
{Spark} M.~K., {O'Donoghue} D., 2015, \mnras, 449, 175

\bibitem[{{Szkody} {et~al}\mbox{.}(2009){Szkody}, {Anderson}, {Hayden},
  {Kronberg}, {McGurk}, {Riecken}, {Schmidt}, {West}, {G{\"a}nsicke}, {Nebot
  Gomez-Moran}, {Schneider}, {Schreiber}, \& {Schwope}}]{szkody09}
{Szkody} P. {et~al.}, 2009, \aj, 137, 4011

\bibitem[{{Townsley} \& {Bildsten}(2003)}]{townsleybildsten03}
{Townsley} D.~M., {Bildsten} L., 2003, \apjl, 596, L227

\bibitem[{{Townsley} \& {G{\"a}nsicke}(2009)}]{townsleygansicke09}
{Townsley} D.~M., {G{\"a}nsicke} B.~T., 2009, \apj, 693, 1007

\bibitem[{{Tremblay} {et~al}\mbox{.}(2016){Tremblay}, {Cummings}, {Kalirai},
  {G{\"a}nsicke}, {Gentile-Fusillo}, \& {Raddi}}]{tremblay16}
{Tremblay} P.-E., {Cummings} J., {Kalirai} J.~S., {G{\"a}nsicke} B.~T.,
  {Gentile-Fusillo} N., {Raddi} R., 2016, \mnras, 461, 2100

\bibitem[{{Uthas} {et~al}\mbox{.}(2011){Uthas}, {Knigge}, {Long}, {Patterson},
  \& {Thorstensen}}]{uthas11}
{Uthas} H., {Knigge} C., {Long} K.~S., {Patterson} J., {Thorstensen} J., 2011,
  \mnras, 414, L85

\bibitem[{{Warner}(1995)}]{warner95}
{Warner} B., 1995, {Cataclysmic Variable Stars}. Cambridge Univ. Press,
  Cambridge

\bibitem[{{Wood} {et~al}\mbox{.}(1986){Wood}, {Horne}, {Berriman}, {Wade},
  {O'Donoghue}, \& {Warner}}]{wood86}
{Wood} J., {Horne} K., {Berriman} G., {Wade} R., {O'Donoghue} D., {Warner} B.,
  1986, \mnras, 219, 629

\bibitem[{{Wood}(1995)}]{wood95}
{Wood} M.~A., 1995, in Lecture Notes in Physics, Berlin Springer Verlag, Vol.
  443, White Dwarfs, {Koester} D., {Werner} K., eds., p.~41

\bibitem[{{York} {et~al}\mbox{.}(2000){York}, {Adelman}, {Anderson},
  {Anderson}, {Annis}, {Bahcall}, {Bakken}, {Barkhouser}, {Bastian}, {Berman},
  {Boroski}, {Bracker}, {Briegel}, {Briggs}, {Brinkmann}, {Brunner}, {Burles},
  {Carey}, {Carr}, {Castander}, {Chen}, {Colestock}, {Connolly}, {Crocker},
  {Csabai}, {Czarapata}, {Davis}, {Doi}, {Dombeck}, {Eisenstein}, {Ellman},
  {Elms}, {Evans}, {Fan}, {Federwitz}, {Fiscelli}, {Friedman}, {Frieman},
  {Fukugita}, {Gillespie}, {Gunn}, {Gurbani}, {de Haas}, {Haldeman}, {Harris},
  {Hayes}, {Heckman}, {Hennessy}, {Hindsley}, {Holm}, {Holmgren}, {Huang},
  {Hull}, {Husby}, {Ichikawa}, {Ichikawa}, {Ivezi{\'c}}, {Kent}, {Kim},
  {Kinney}, {Klaene}, {Kleinman}, {Kleinman}, {Knapp}, {Korienek}, {Kron},
  {Kunszt}, {Lamb}, {Lee}, {Leger}, {Limmongkol}, {Lindenmeyer}, {Long},
  {Loomis}, {Loveday}, {Lucinio}, {Lupton}, {MacKinnon}, {Mannery}, {Mantsch},
  {Margon}, {McGehee}, {McKay}, {Meiksin}, {Merelli}, {Monet}, {Munn},
  {Narayanan}, {Nash}, {Neilsen}, {Neswold}, {Newberg}, {Nichol}, {Nicinski},
  {Nonino}, {Okada}, {Okamura}, {Ostriker}, {Owen}, {Pauls}, {Peoples},
  {Peterson}, {Petravick}, {Pier}, {Pope}, {Pordes}, {Prosapio},
  {Rechenmacher}, {Quinn}, {Richards}, {Richmond}, {Rivetta}, {Rockosi},
  {Ruthmansdorfer}, {Sandford}, {Schlegel}, {Schneider}, {Sekiguchi}, {Sergey},
  {Shimasaku}, {Siegmund}, {Smee}, {Smith}, {Snedden}, {Stone}, {Stoughton},
  {Strauss}, {Stubbs}, {SubbaRao}, {Szalay}, {Szapudi}, {Szokoly}, {Thakar},
  {Tremonti}, {Tucker}, {Uomoto}, {Vanden Berk}, {Vogeley}, {Waddell}, {Wang},
  {Watanabe}, {Weinberg}, {Yanny}, {Yasuda}, \& {SDSS Collaboration}}]{york00}
{York} D.~G. {et~al.}, 2000, \aj, 120, 1579

\bibitem[{{Zorotovic}, {Schreiber} \& {G{\"a}nsicke}(2011){Zorotovic},
  {Schreiber}, \& {G{\"a}nsicke}}]{zorotovic11}
{Zorotovic} M., {Schreiber} M.~R., {G{\"a}nsicke} B.~T., 2011, \aap, 536, A42

\end{thebibliography}

\end{document}